# Simulating Brain Tumor Heterogeneity with a Multiscale Agent-Based Model:

## Linking Molecular Signatures, Phenotypes and Expansion Rate

**Le Zhang, Costas G. Strouthos [#], Zhihui Wang, and Thomas S. Deisboeck ***

Complex Biosystems Modeling Laboratory, Harvard-MIT (HST) Athinoula A. Martinos Center for Biomedical Imaging, Massachusetts General Hospital, Charlestown, MA 02129, USA.

**\*Corresponding Author:**

Thomas S. Deisboeck, M.D.
Complex Biosystems Modeling Laboratory
Harvard-MIT (HST) Athinoula A. Martinos Center for Biomedical Imaging
Massachusetts General Hospital-East, 2301
Bldg. 149, 13th Street
Charlestown, MA 02129
Tel: 617-724-1845
Fax: 617-726-7422
Email: deisboec@helix.mgh.harvard.edu

[#] Current address: École Polytechnique Fédérale de Lausanne (EPFL), BCH 3121, CH-1015 Lausanne. Email: costas.strouthos@epfl.ch

**ABSTRACT**

We have extended our previously developed 3D multi-scale agent-based brain tumor model to simulate cancer heterogeneity and to analyze its impact across the scales of interest. While our algorithm continues to employ an epidermal growth factor receptor (EGFR) gene-protein interaction network to determine the cells' phenotype, it now adds an explicit treatment of tumor cell adhesion related to the model's biochemical microenvironment. We simulate a simplified tumor progression pathway that leads to the emergence of five distinct glioma cell clones with different EGFR density and cell 'search precisions'. The *in silico* results show that microscopic tumor heterogeneity can impact the tumor system's multicellular growth patterns. Our findings further confirm that EGFR density results in the more aggressive clonal populations switching earlier from proliferation-dominated to a more migratory phenotype. Moreover, analyzing the dynamic molecular profile that triggers the phenotypic switch between proliferation and migration, our *in silico* oncogenomics data display spatial and temporal diversity in documenting the regional impact of tumorigenesis, and thus support the added value of multi-site and repeated assessments *in vitro* and *in vivo*. Potential implications from this *in silico* work for experimental and computational studies are discussed.

**1. INTRODUCTION**

We present the extension of our previously developed three-dimensional (3D) *agent-based*, multi-scale brain tumor model to incorporate a simplified glioma progression pathway in an effort to study the impact of clonal heterogeneity on tumor growth dynamics. For the micro-macroscopic environment, we expanded the initial setup [1-3] by simulating the interaction between the chemoattractant concentration and the degree of cell adhesion. Like before [3], a set of chemoattractants such as glucose, oxygen tension and transforming growth factor alpha ($TGF_\alpha$) diffuses [4-6] along a gradient that depends both on the initial conditions as well as on the cells' consumption and secretion ($TGF_\alpha$). Also, the glioma cells' considerably high migration rate *in vivo*, and even more so, *in vitro* [7] is impacted by the $TGF_\alpha$ concentration to properly reflect available experimental data [8-11]. Specifically, we modeled the chemotactically attracted cells as follows: once the concentration of the so called homotype attractant [12], in this case $TGF_\alpha$, substantially exceeds that of the heterotype attractant [13] such as glucose, the homotype attractant ($TGF_\alpha$) plays a much more important role for directing cell attraction and vice versa. For the molecular environment, each cell is equipped with an epidermal growth factor receptor (EGFR) *gene-protein interaction pathway* [1-3] that is implicitly integrated with a simplified *cell cycle* module [14]. The pathway molecule 'activated phospholipase Cγ' (*PLCγ*) is critical in determining the cell phenotype which represents experimental results reported in part by [15] for the case of breast cancer. However, other than in our previous works where heterogeneity arose only from phenotypic alterations of an otherwise mono-clonal tumor, here, we have integrated a simplified *cancer progression* model (based on [16]) to generate a heterogeneous brain tumor that eventually consists of five distinct clonal populations that emerged through mutational events. These clones differ in their EGFR density in addition to being able to adjust their phenotypic behavior readily to the micro-environmental conditions. To properly reflect available experimental data, the sub-clones with higher EGF receptor density are more aggressive [17], and operate with an inefficient metabolism [18, 19]; yet, they achieve a higher chemotactic search precision to move faster along the environment's least resistance, most permission and highest attraction paths [20]. We will discuss the results and their potential implications, and introduce plans for future work.

## 2. PREVIOUS WORKS

Earlier tumor models focused primarily *either* on the micro-macroscopic *or* on the molecular level to simulate tumor growth and migration. For instance, for the micro-macroscopic scale, Mansury et al. [21-23] proposed a two-dimensional (2D) agent-based model in which the spatio-temporal expansion of malignant brain tumor cells is guided by environmental heterogeneities (mechanical confinement, toxic metabolites and nutrient sources) in order to gain more insight into the systemic effect of such cellular chemotactic search precision modulations. For the molecular scale, Schoeberl et al. [24] built up a computational model to describe the dynamics of the MAPK cascade activated by EGF receptors and Araujo et al. [25] developed an EGFR signaling network model that not only shows the protein dynamics in the EGFR pathway but also predicts the therapeutic effect of inhibitors on the molecular network. In an effort to develop a cell cycle module, Alarcon et al. [14] extended the work of Tyson and Novak [26] to model the cancer cell cycle process under hypoxic conditions. Additionally, to simulate mutation among different clones, Sole and Deisboeck [16] revised the previous quasispecies model of Swetina and Schuster [27] to research the impact of genetic instability on tumor growth. While these models undoubtedly are useful at their particular scale of interest, the quest of studying the tumor as a biological system that crosses scales requires a *multiscale* modeling approach. Therefore, in [28] and [1, 2] our group developed a 2D multi-scale agent-based tumor model to integrate both, the molecular and micro-macroscopic environments. Specifically, in [1] we put forward an EGFR gene-protein interaction pathway in which activated *PLCγ* is employed to determine the cell phenotype. Subsequently, in [2] we then compared the biological behavior of glioma cell clones with varying EGFR density. Building on this work, in [3] we recently introduced a 3D multi-scale agent-based brain tumor model which explicitly includes a cell cycle description. However, this multi-scale model has a number of shortcomings that include simulating a monoclonal, and thus a genetically stable tumor, and it also suffers from a rather simplistic description of the tissue microenvironment. In an effort to model these features more realistically the subsequent section will describe the setup of a much more sophisticated 3D multi-scale brain tumor model that includes explicitly the phenotype of *cell adhesion* as it relates to chemoattractant [10] and incorporates specifically the emergence of *heterogeneous* tumor cell clones.

## 3. MATHEMATICAL MODEL

As before, this multi-scale agent-based brain tumor model includes macro-microscopic and molecular environments that interact with each other. The following sections describe these environments and their relationships in detail.

### 3.1. Macro-microscopic environment

### 3.1.1. Clonal heterogeneity

Here, a slice of virtual brain tissue is modeled with a 100*100*100 3D rectangular lattice as depicted in **Figure 1**. A continually replenished nutrient source, representing a blood vessel, is located in Cube 4.

**Figure 1**

To yield tumor heterogeneity, an element of *genetic instability* is introduced that is based on the mutation model presented by [16]. Our 'progression pathway' here incorporates five glioma cell clones (A, B, C, D and E) that emerge sequentially (**Figure 2**), and that are distinct in their EGFR receptor density. Based on experimental data, in our setup, cell clones with higher EGFR receptor density are more aggressive [17, 29, 30] and operate with an inefficient metabolism [18, 19]; yet, they achieve a higher chemotactic search precision to move faster along the environment's least resistance, most permission and highest attraction paths [20].

**Figure 2**

In the beginning, five hundred clone A cells are initialized as close as possible to the center of the lattice as each cell occupies one lattice point. As a first iteration, the model implements a *linear* progression pathway in assuming that clone A can only mutate to B, clone B to C, C to D, and D to E. The respective mutation rates depend on the cell's proliferation period and are listed in **Table 1**.

**Table 1**

To mimic progression from a relatively genetically stable precursor, we set the mutation rate from clone A to B as $1 - \dfrac{\text{B Proliferation Period}}{\text{A Proliferation Period}} = 10\%$. Applying the same calculation for the remaining clones, we deduce a mutation rate from B to C of 11%, from C to D of 12% and D to E of 15%.

### 3.1.2. Chemoattractants

Transforming growth factor alpha ($TGF_\alpha$) is a protein that not only can affect the secreting tumor cell itself, hence generates an 'autocrine' feedback loop, but also can impact bystander cells, an effect known as 'paracrine'. Habib et al. [12] refers to this soluble chemical effector as homotype chemoattractant. Conversely, glucose, not produced by tumor cells, is denoted as heterotype attractants [13]. Our setup places a blood vessel in the center of Cube 4 to supply glucose, $TGF_\alpha$ and oxygen for the developing brain tumor. Initially, these chemoattractants are dispersed by normal distribution described in **Eq. 1-3**:

$$X_1^{ijk} = T_m \cdot \exp(-2d_{ijk}^2 / \sigma_t^2) \tag{1}$$

$$X_{14}^{ijk} = G_a + (G_m - G_a) \cdot \exp(-2d_{ijk}^2 / \sigma_g^2) \tag{2}$$

$$k_{44}^{ijk} = k_a + (k_m - k_a) \cdot \exp(-2d_{ijk}^2 / \sigma_o^2) \tag{3}$$

In **Eq. 1**, $T_m$ stands for the maximum $TGF_\alpha$ ($X_1$) concentration in the tumor [31] and $\sigma_t$ is the parameter that controls the dispersion of the $TGF_\alpha$ level. In **Eq. 2**, $G_a$ is the minimum blood glucose ($X_{14}$) level while $G_m$ stands for the maximum concentration of glucose in blood, with $\sigma_g$ being the parameter controlling the dispersion of glucose. Likewise, in **Eq. 3**, $k_a$ is the minimum oxygen tension and $k_m$ represents the maximum oxygen tension ($k_{44}$) [3, 14], $\sigma_o$ is the parameter controlling the dispersion of the oxygen tension level, and $d_{ijk}$ is the L-infinity [21] distance to the center of Cube 4. During the simulation, at each time step the cell will take up glucose to maintain its metabolism and secrete $TGF_\alpha$ [1] triggering both paracrine and autocrine biological behavior that is described by **Eq. 4-5**:

$$X_1^t = X_1^{t-1} + S_T \tag{4}$$

$$X_{14}^t = X_{14}^{t-1} - r_n, n = A, B, C, D, E \tag{5}$$

In **Eq. 4-5**, $t$ represents the time step, $S_T$ is the $TGF_\alpha$ secretion rate [32] and $r_n$ is the cell's glucose uptake coefficient [33] for each clone. Because [19] argues that the metabolism of cancer cells is approximately eight times larger than the metabolism of normal cells and since [33] discovered that for U87 human glioma cells the glucose uptake rate is 0.77 pmol/h per cell, we assign each clone (A, B, C, D and E) a different glucose uptake rate related to its EGFR density, which is listed in the **Table 2**.

**Table 2**

Because the chemo-attractants' initialization through **Eq. 1-3** causes different gradients in the 3D lattice, the aforementioned biochemicals glucose, $TGF_\alpha$ and oxygen will simultaneously diffuse [1-3] during the simulation according to **Eq.6-8**.

$$\frac{\partial X_1^{ijk}}{\partial t} = D_2 \cdot \nabla^2 X_1^{ijk}, t = 1,2,3... \tag{6}$$

$$\frac{\partial X_{14}^{ijk}}{\partial t} = D_1 \cdot \nabla^2 X_{14}^{ijk}, t = 1,2,3... \tag{7}$$

$$\frac{\partial k_{44}^{ijk}}{\partial t} = D_o \cdot \nabla^2 k_{44}^{ijk}, t = 1,2,3... \tag{8}$$

where $D_1$ is the diffusion coefficient of glucose [5], $D_2$ is the $TGF_\alpha$ diffusion coefficient [6] and $D_o$ stands for the diffusion coefficient of oxygen [4].

### 3.1.3. Spatial search process

Representing the biological mechanism of *chemotaxis* we have again implemented a search mechanism as put forward already in [21, 22].

$$T_{ijk} = \psi_n \cdot L_{ijk} + (1 - \psi_n) \cdot \varepsilon_{ijk}, n = A, B, C, D, E \tag{9}$$

where $T_{ijk}$ stands for the perceived attractiveness of location $(i,j,k)$, $L_{ijk}$ represents the correct, non-erroneous evaluation of location $(i,j,k)$ where $\varepsilon_j \sim N(\mu, \sigma^2)$ is an error term that is normally distributed with mean $\mu$ and variance $\sigma^2$. The parameter $\psi_n$ is positive between zero and one, $0 \leq \psi_n \leq 1$ , and represents the extent of the 'search precision' (with '0' exhibiting purely *random walk* behavior, and '1' being fully *biased* without any error in signal processing); the subscript of $\psi$ represents the different clones. In accordance with experimental data [29] that report an acceleration of the tumor's spatio-temporal expansion rate with increasing EGFR density per cell, we argue that a cancer cell with high EGFR density should express a comparably high search precision (being lesser affected by the noise function). Thus, we assign a search precision value of 0.5 for clone A and B, and 0.7 for clone C, D and E, respectively.

The definitions of the Von Neumann and Moore neighborhoods [21] usually are restricted to 2D lattices. We therefore define 3D Von Neumann neighborhood, 3D Diagonal neighborhood, and 3D Moore neighborhood as detailed in **Eq.10-12** with

$$N^v_{(x_0,y_0,z_0)} = \left\{ \begin{array}{l} (x,y,z) : 0 < |x-x_0| + |y-y_0| + |z-z_0| \leq 1 \\ \text{and } |x-x_0| = 1 \text{ or } 0, |y-y_0| = 1 \text{ or } 0, \ |z-z_0| = 1 \text{ or } 0 \end{array} \right\} \tag{10}$$

$$N^d_{(x_0,y_0,z_0)} = \left\{ \begin{array}{l} (x,y,z) : 0 < |x-x_0| + |y-y_0| + |z-z_0| \leq 2 \\ \text{and } |x-x_0| = 1 \text{ or } 0, |y-y_0| = 1 \text{ or } 0, \ |z-z_0| = 1 \text{ or } 0 \end{array} \right\} \tag{11}$$

$$N^m_{(x_0,y_0,z_0)} = \left\{ \begin{array}{l} (x,y,z) : 0 < |x-x_0| + |y-y_0| + |z-z_0| \leq 3 \\ \text{and } |x-x_0| = 1 \text{ or } 0, |y-y_0| = 1 \text{ or } 0, \ |z-z_0| = 1 \text{ or } 0 \end{array} \right\} \tag{12}$$

Here, ($x_0$ , $y_0$ , $z_0$) denotes the current location of the cell and ($x$ ,$y$ ,$z$) represents the coordinates of its neighborhood. We set the grid size to 10 μm to reflect the (idealized) average diameter of a real tumor cell and each simulation step represents approximately 2.5 hours, similar to our previous study [3]. The maximum distance for a given tumor cell to migrate under a 3D Von Neumann neighborhood search is therefore 10 μm/simulation step, 14.4 μm/simulation step under a 3D Diagonal neighborhood search, and 17.3 μm/simulation step under a 3D Moore neighborhood search paradigm, hence well within the range of the tumor cell migration rate reported *in vitro* [7].

It has been well established that cancer cells differ from normal cells in a variety of ways which include uncontrolled proliferation, loss of contact inhibition, progressive lack of tumor cell-tumor cell adhesion and tissue invasion [34]. Also, Athale and Deisboeck [2] already demonstrated in separate simulation runs that *in silico,* glioma cells with higher EGFR density are more aggressive with faster migration and lower proliferation rate. Moreover, as indicated by [8-11] the concentration of chemo-attractant can impact the degree of tumor cell adhesion and the cells' migration rate. We therefore ranked the diffusive $TGF_\alpha$ concentration at three levels in decreasing order: If the neighborhood of the tumor cell carries a high average $TGF_\alpha$ concentration (= level 1) the cell can choose its next location from the 3D Moore neighborhood ( $N_{(x_0, y_0, z_0)}^m$ ) which conveys the highest spatial permission and therefore results in the fastest cell migration rate. If the cell resides at a location within the second $TGF_\alpha$ concentration level (2), it will choose the next location from the 3D Diagonal neighborhood ( $N_{(x_0, y_0, z_0)}^d$ ). Otherwise, the cell will select its next location from the more restrictive 3D Von Neumann neighborhood ( $N_{(x_0, y_0, z_0)}^v$ ). We therefore incorporate in our expanded model here not only that the chemotactically-driven search precision of clones C, D, E is higher than that of clones A and B. Rather, in addition, we implement that a higher $TGF_\alpha$ concentration can increase the cell's spatial "permission", thus reflecting its progressive loss of tumor cell-tumor cell adhesion as reported in [8-11], and as such implicitly adding an element of *haptotaxis*.

As discussed earlier, $L_{ijk}$ is the correct, non-erroneous evaluation of location, which includes homotype and heterotype chemoattractants. **Eq. 13** models this biological behavior, i.e., if a cell resides in location ($x_0$, $y_0$, $z_0$) and the sum of its 3D Von Neumann neighborhood $TGF_\alpha$ ( $\sum_{N_{(x_0, y_0, z_0)}^v} X_1$ ) is greater than glucose concentration ( $\sum_{N_{(x_0, y_0, z_0)}^v} X_{14}$ ), then this cell is subjective to choose the homotype attractant, otherwise the cell is attracted by the heterotype cue.

$$L_{ijk} = X_1^{ijk}, \text{if } \frac{\sum_{N_{(x_0,y_0,z_0)}^V} X_1}{X_1} > \frac{\sum_{N_{(x_0,y_0,z_0)}^V} X_{14}}{X_{14}} \tag{13a}$$

$$L_{ijk} = X_{14}^{ijk}, \text{if } \frac{\sum_{N_{(x_0,y_0,z_0)}^V} X_1}{X_1} < \frac{\sum_{N_{(x_0,y_0,z_0)}^V} X_{14}}{X_{14}} \tag{13b}$$

### 3.1.4. Cell phenotype

As before, we simulate four different cellular phenotypes, i.e. migration, proliferation, quiescence and apoptosis – now, however, for each of the five different glioma cell clones. We again implement the concept of 'dichotomy' for glioma cells, first put forward by [35] which argues on the basis of mounting experimental data that migration and proliferation are mutually exclusive behaviors that can happen in sequence, yet hardly at the same time in the same cell. Using breast cancer cells, Dittmar et al. [15] discovered that *PLCγ* is activated transiently during cell migration. Building on this experimental data, Athale et al. [1, 2] and Zhang et al. [3] employ a *PLCγ* threshold ($\sigma_{PLC}$) to model a hypothesized biological *switching* behavior in that *PLCγ* is activated to a greater extent during migration and more gradually during proliferation. Thus, the *migration potential* is assessed by a cell evaluating the following function:

$$M_n[(PLC\gamma)] = [\frac{d(PLC\gamma)}{dt}] \tag{14}$$

where $M_n$ is the migration potential that equals the change in concentration of active *PLCγ* over time and $n$ is the cell number. Once $M_n$ is greater than $\sigma_{PLC}$, the cell chooses a migratory phenotype, otherwise it undergoes proliferation. We note that there are three possibilities that the cell enters the reversible quiescent state: (1) the cell is unable to find an empty location to migrate to or proliferate into; or (2) the migration potential ($M_n$) is less than $\sigma_{PLC}$ and $ppTGF\alpha - EGFR\_s$ is less than $\sigma_{EGFR}$; or, (3) the glucose concentration around the cell drops to critically low values. If the on-site glucose concentration diminishes even further, the cell enters an apoptotic state [36].

### 3.2. Molecular environment

The sub-cellular setup is described in detail in [3]. Here, we therefore only briefly summarize this molecular environment, focus on its modifications, and note that the model's intracellular workflow is depicted in **Figure 3.(a)**. Specifically, each cell is equipped with an *EGFR gene-protein interaction network* shown in **Figure 3.(b)**, which is composed of 14 molecular species in addition to the five molecular species involved in our simplified *cell cycle* representation as listed in **Table 3**. These molecular species interact with each other according to mass balance equations, denoted by the general form:

$$\frac{dX_i}{dt} = \alpha_i X_i - \beta_i X_i \tag{15}$$

where $X_i$ is the mass of *ith (i=1-14)* molecules comprising the EGFR network and $\alpha_i, \beta_i$ *(i=1-14)* are the rate of synthesis and degradation rate of EGFR pathway, respectively. Also, $X_i$ *(i=15-19)* molecules are the components of the cell cycle and $\alpha_i, \beta_i$ *(i=15-19)* are the coefficients of the cell cycle module. The values of parameters and constants are listed in [3]. As an important extension of our previous works, a proliferation process has now been *explicitly* integrated into the model in that more aggressive cell clones are associated with a shorter *cell doubling time*.

**Figure 3**

**Table 3**

## 4. RESULTS

Our code is implemented in Java (Sun Microsystems, Inc., USA), combined with in-house developed classes for representing molecules, reactions and multi-receptors as a set of hierarchical objects. Running the simulation 10 times with different random normal distribution, $\sigma_g$, of glucose (**Eq. 2**), the algorithm requires a total of 70 hrs 46 min of CPU time on a computer with an IBM Bladecenter machine (dual-processor 32-bit Xeons ranging from 2.8-3.2GHz w/2.5GB RAM) and Gigabit Ethernet. Each node runs Linux with a 2.6 kernel and Sun's J2EE version 1.5.

Employing a multi-level growth model, our analysis spans several scales of interest in starting with the overall, macroscopic behavior followed by the multicellular population patterns before focusing on the distinct impact molecular-level dynamics have on microscopic cell phenotypes.

**Volumetric tumor growth dynamics:** We measure the tumor system's [total] volume by counting the number of lattice sites occupied by a tumor cell regardless of its phenotype, hence lumping together both proliferative- and migratory-driven expansion. **Figure 4** shows the tumor's volumetric increase over time for the 10 simulation runs.

**Figure 4**

To achieve a higher spatial resolution of analysis, we divide the macroscopic tumor into three microscopic regions, sectioning the distance between the tumor and the nutrient source such as shown in **Figure 5.(a)**: region 1 is closest to the nutrient source, region 3 is farthest from the nutrient source, with region 2 being located in between.

**Tumor heterogeneity:** In an attempt to quantify clonal heterogeneity we employ the so-called *Shannon Index* which is defined as $H = -\sum_i p_i \ln(p_i)$ where $p_i$ is the frequency of clone $i$ within the tumor [37]. **Figure 5.(b)** displays this index for region 1, 2, and 3 at each time step. While the indices for all three regions generally increase (i.e., exhibit more clonal heterogeneity) region 1 experiences an unexpected drop around $t = 100$, which indicates the emergence of a more homogenous sub-section prior to a gain of clonal diversity similar to that of the other two geographic sub-regions.

**Figure 5**

Visualizing this type of emergent heterogeneity by orienting us on curve-crossing points seen in **Figure 5.(b)**, we display in **Figure 6.(a)-(c)** representative 3D snapshots of the tumor at consecutive time points $t = 105$, 180 and 230. As one would expect from a chemotaxis-driven search, the tumor overall expands towards the location of the nutrient source (compare with **Figure 1**).

**Figure 6**

**Expansion rate per tumor region:** To support this impression quantitatively, we measure the expansion rate for each sub-region by computing the average expansion radius for each region based on its center of mass. The results show that tumor region 1, i.e., closest to the nutrient source, expands much faster than regions 2 and 3 that are farther from the nutrient source (**Figure 7**). The lesser expansion rate of region 2 (versus region 3) is due to the fact that cells in region 2 find fewer empty locations to choose from.

**Figure 7**

**Expansion rate per tumor clone:** The next question to address is then how much of that regional specificity is due to its prevailing clonal population and as such we measure the expansion rate for clone A, B, C, D and E at each time step by computing the average expansion radius for each clone based on the center of each clone. From **Figure 8** one can see that clone E (with the highest EGFR density per cell) exhibits the highest expansion rate increase. In contrast, although clone A and B (with lower EGFR density) originate first, and thus also start to expand much earlier than clone C, D and E, these two lower malignant clones keep a much smaller expansion rate for most of the simulation.

So far, the results support the notion that a higher density of the more aggressive clone E is responsible for the higher expansion rate increase in the tumor region that is closest to the nutrient source. This therefore warrants a more detailed look into the phenotypic dynamics of the cancerous clones in general, and of clone E in particular.

**Figure 8**

**Phenotypic spectrum per tumor clone:** Turning then to the microscopic analysis level by monitoring the phenotypic fractions of clone A, B, C, D and E at each time step (**Figure 9. (a)-(e)**) it is evident that clones with higher EGFR density are eventually comprised of a larger migratory fraction, while exhibiting a smaller proliferation and quiescence cell population. Specifically, clone E has the largest and most sustained migratory cell population, and, as the curve shows, the 'population' crossover between a more stationary, proliferation-dominated to an expansive, migration-dominated clone happens earlier than in the other, less aggressive tumor clones. We also note that, qualitatively, the fluctuations in the curves increase from a fairly periodic pattern in the lower malignant clones A and B, to a much less regular, 'noisy' pattern in the higher malignant clones.

**Figure 9**

**Figure 10**

**Molecular phenotype-switching profiles:** Moving now to the sub-cellular level, and building on our previous study [3], we use a heatmap display to visualize the percentage change of each molecular component of the implemented EGFR network (**Figure 3. (b)**) when the cells switch their phenotype from migration to proliferation or from proliferation to migration. We focus on time points, $t$ =200 and 230, i.e. when the simulation begins to show sustained migration for most of the clones and again, when the run is almost completed. From **Figure 10** it is evident that no phenotype switching signature is exactly alike. That is, differences exists between time points, phenotype switches, and geographic regions. However, there are also some similarities as for most cases, the signals for dimeric $TGF\alpha - EGFR$ cell surface complex ( $X_3$ ), phosphorylated active dimeric $TGF\alpha - EGFR$ cell surface complex ( $X_4$ ), cytoplasmic inactive dimeric $TGF\alpha - EGFR$ complex ( $X_5$ ), cytoplasmic $EGFR$ protein ( $X_6$ ) and Ca-bound inactive $PLC_\gamma$ ( $X_{10}$ ) experience a more profound dynamic change for each region.

## 5. DISCUSSION & CONCLUSIONS

In here, we present an expansion of our 3D multi-scale agent-based brain tumor model that has now been extended to (**1**) also incorporate an element of *genetic instability* to simulate glioma progression and to analyze its impact across multiple scales. Building on our previous works [1-3], this model not only encompasses molecular, microscopic and macroscopic scales of observation, but now also integrates (**2**) both, *cell doubling time* and cell cycle into the cells' proliferation module. (**3**) This model also simulates the impact of chemo-attractants on cancer *cells adhesion* and migration rate. In the following, we discuss the results in more detail.

The tumor region adjacent to a blood vessel representing nutrient source (i.e., region 1) harbors early on the highest clonal heterogeneity (**Figure 5.(b)**), reflecting the spatio-temporal output of the implemented linear progression path. However, around time step $t = 100$ this geographic 'tip' region (**Figure 6.(a)-(c)**) becomes increasingly homogeneous due to a temporary *competitive* advantage of the most aggressive clone E with regards to its chemo-sensing ability and high spatial expansion rate. During a period that is marked by variability across runs (increasing fluctuations in standard deviation), clone E prevails in its quest to move fastest into the direction of "(least resistance,) most permission and highest attraction" [20], and, in the process, impacting the spatio-temporal patterns of the entire tumor system. However, the micro-environmental conditions closer to the core of the tumor, i.e. for cells in regions 2 and 3 are less selective for clone E and thus continuous tumor progression yields a sustained increase in the *Shannon Index* that eventually even exceeds the peak value for region 1. One can argue that, other than region 1, regions 2 and 3 allow for an element of temporary *coexistence* if not *cooperation* between the clones. Since regions 2 and 3 together add up to a larger volume than region 1 alone, this finding corresponds well with another recent claim of cancer cell cooperation fostering tumorigenesis [38], and also with the results reported by [37] on esophagus cancer which indicated that clonal heterogeneity is a greater risk factor for progression to cancer than the expansion of a homogeneous clone with a low *Shannon Index* value. This is confirmed in our own results as clone E's dominance of region 1 is relatively short-lived with the heterogeneity index soon returning to values comparable to the two proximal tumor regions. The question arises then as to what precisely are the effects of such clonal heterogeneity on tumor growth dynamics? First, due to their higher density of EGF receptors per cell, clones C, D and primarily E exhibit an increased expansion rate (**Figure 8**) over clones A and B. Correspondingly, **Figure 9** confirms that these tumor clones with higher EGFR density switch earlier to an invasive phenotype-dominated cell population that operates with a larger fraction of migratory cells, yet with lesser proliferative and quiescent cells. Our *in silico* model shows that higher EGFR expression leads to faster expansion of tumor areas that harbor more such aggressive cell populations, in areas of nutrient abundance. This, in turn, will result in overall growth asymmetries (see region 1, **Figure 7**), i.e., findings that taken together correspond well with reported experimental and clinical results [17, 29, 30].

Taking advantage of our algorithm's multi-scale design we can analyze the molecular composition of the gene-protein interaction network at the location and time when the cell transitions from one phenotype to another. Specifically, **Figure 10** demonstrates that there are both spatially *and* temporally distinct *molecular signatures* that lead to the cells' phenotypic switch. Cautiously extrapolated to an orders of magnitude more complex experimental situation, one wonders how much *more* robust *in vitro* and *in vivo* expression profiles would have to be within a patient's tumor (let alone across an entire patient population) to warrant the single site and time point assessments that are commonly used for functional genomics in clinics. In fact, applying a global genomic analysis Maher et al. [39] recently showed widespread differences between primary and secondary glioblastoma entities and reported that the latter group (which develops over multiple years from lower malignant astrocytomas [40]) is heterogeneous in its molecular pathogenesis. Our findings should therefore not be misinterpreted as questioning the value of such high-throughput data, rather, our *in silico* results argue that assessing the molecular profile *repeatedly*, and at *different* sites of the tumor, should *add* predictive power. Interestingly, we found that at time step 200 the percentage changes of some molecules (i.e., $X_3$, $X_4$, $X_5$, $X_6$ and $X_{10}$) in region 2 and 3 is more substantial than in region 1 regardless of the switch *per se*. The reason for this is that the higher EGFR density in clones C-E results in higher concentrations of other downstream molecules as well. Hence, the relative molecular fluctuations should be lesser in clones C, D and E than in clones A and B. And, since region 1 is mostly comprised of cells derived from clones C-E, its molecular composite signature appears to be more preserved between the time points than that of the other two regions. However, over time many more aggressive cells originate from clones A and B also in the more proximal tumor regions, and we see therefore eventually a somewhat more homogeneously aggressive pattern where regions 2 and 3 start to resemble the distant region 1. As such, by monitoring the molecular signature dynamics per region, one can at least *in silico* visualize the gradual tumorigenic evolution of sub-regions of the tumor, in an effort to better understand how they may contribute to multicellular growth behavior.

Despite these considerable extensions, our modeling platform of course has still a number of shortcomings that can and should be addressed in future works. For example, we currently rely on a fixed mutation rate between each clone and mutations only occur linearly from A to E, without any non-linear 'shortcuts' and without assigning different weight to specific mutations. In subsequent research, we intend to implement a dynamical mutation rate that depends both on intrinsic and extrinsic or microenvironmental factors. Other extensions on the molecular level will rely on sensitivity analysis, or may include a more explicit treatment of vascular architecture. Nonetheless, this work significantly enhances our previously introduced platform [1-3] by including the element of genetic instability and integrating both cell cycle and cell doubling time into the cells' proliferation process. Given the fact that we begin to be able to validate some of the model's predictions with experimental studies, we firmly believe that this type of interdisciplinary approach holds significant promise to help improve our understanding of brain cancer progression and its impact on the tumor system's regional and global growth dynamics.

## ACKNOWLEDGEMENTS

This work has been supported in part by NIH grants CA 085139 and CA 113004 (The Center for the Development of a Virtual Tumor, CViT at http://www.cvit.org) and by the Harvard-MIT (HST) Athinoula A. Martinos Center for Biomedical Imaging and the Department of Radiology at Massachusetts General Hospital.

## CAPTIONS

**Figure 1.** Schematic of 3D lattice, with Cube 4 harboring the nutrient source.

**Figure 2.** The mutational steps that constitute the genetic progression pathway from clone A to E.

**Figure 3. (a)** The cancer cell's sub-cellular workflow that determines its phenotypic output. **(b)** The EGFR gene-protein interaction network (see text for details).

**Figure 4.** The volume of the tumor system (*y-axis*) over time (*x-axis*). Shown are mean values of 10 simulation runs with random standard deviation (*grey*) of the glucose concentration at a range between 50 and 60 nmol/lattice site.

**Figure 5. (a)** The schematic shows the separation of the tumor into three regions (*dotted lines*) with varying distance to the nutrient source (*black circle* in Cube 4; compare with **Fig. 1**). **(b)** Depicted is the Shannon Index (*y-axis*) over time (*x-axis*) for these three distinct regions. Shown is the result of 10 runs with standard deviation.

**Figure 6.** 3D snapshots of the tumor system at time step *t* = 100 (**a**), *t* =180 (**b**), *t* = 230 (**c**). The *red circle* depicts the location of the nutrient source (compare with **Fig. 1**). Note, the light and dark *green* colors represent cells of clone A in their proliferation state and migration state, respectively; light and dark *blue* represent cells of clone B in proliferation and migration states; light and dark *yellow* represent cells of clone C in proliferation and migration state; light and dark *purple* represent cells of clone D in proliferation and migration states; light and dark *red* represent cells of clone E in proliferation and migration states. Finally, light and dark *grey* represent cells in the quiescence and apoptosis states, respectively, for clones A-E.

**Figure 7.** The expansion rate (*y-axis*) of regions 1, 2 and 3 over time (*x-axis*).

**Figure 8.** The expansion rate (*y-axis*) of clone A, B, C, D and E over time (*x-axis*).

**Figure 9.** The phenotypic spectrum (*y-axis*) for clone (**a**) A, (**b**) B, (**c**) C, (**d**) D and (**e**) E over time (*x-axis)*.

**Figure 10.** Percentage change of each EGFR network component (compare with **Fig. 3(b)**) combined to molecular signatures and represented as heatmaps, for the three geographic tumor regions, at time step 200 and 230, respectively. (Note: M → P = phenotypic switch from migration to proliferation; P → M = from proliferation to migration [3]).

**Table 1.** Cell cycle times and proliferation periods for clone A to E.

**Table 2.** Coefficients of the tumor heterogeneity model taken from the literature [1-3, 5, 6].

**Table 3:** Symbols of the EGFR gene-protein interaction network and cell cycle module taken from the literature[3].

# FIGURES & TABLES

**FIGURE 1.**

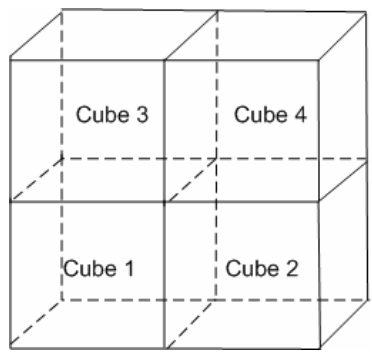

**FIGURE 2.**

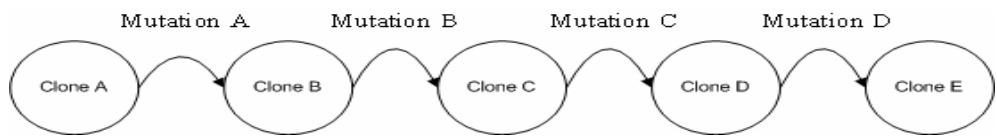

**FIGURE 3. (a)**

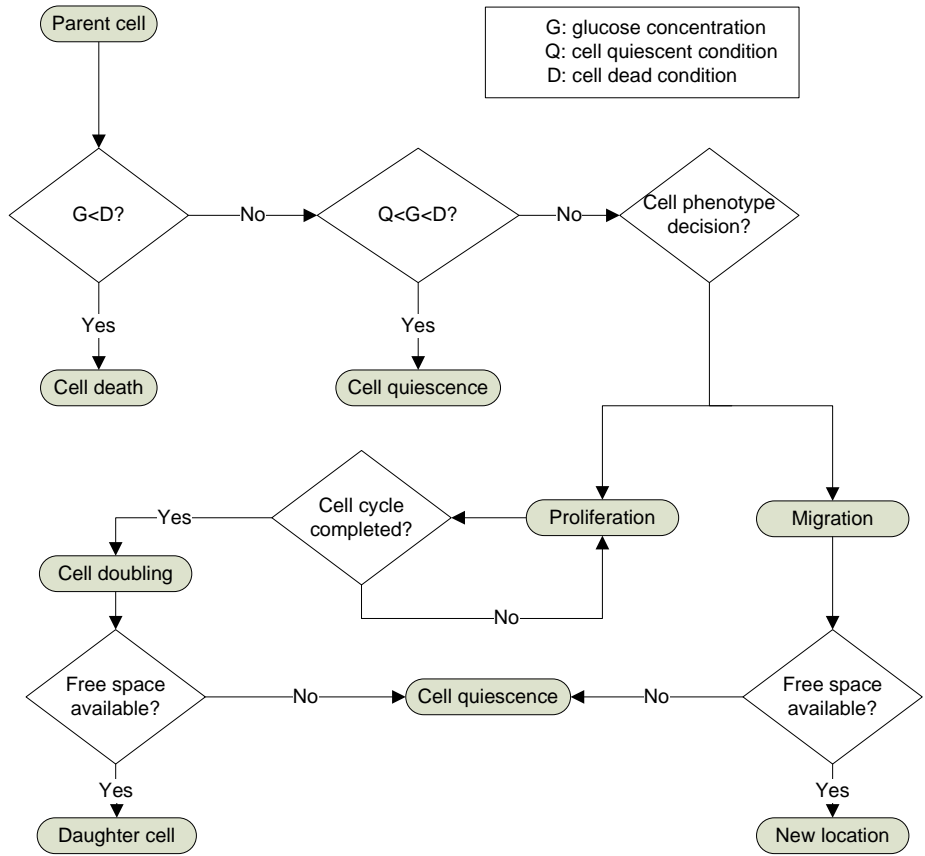

**(b)**

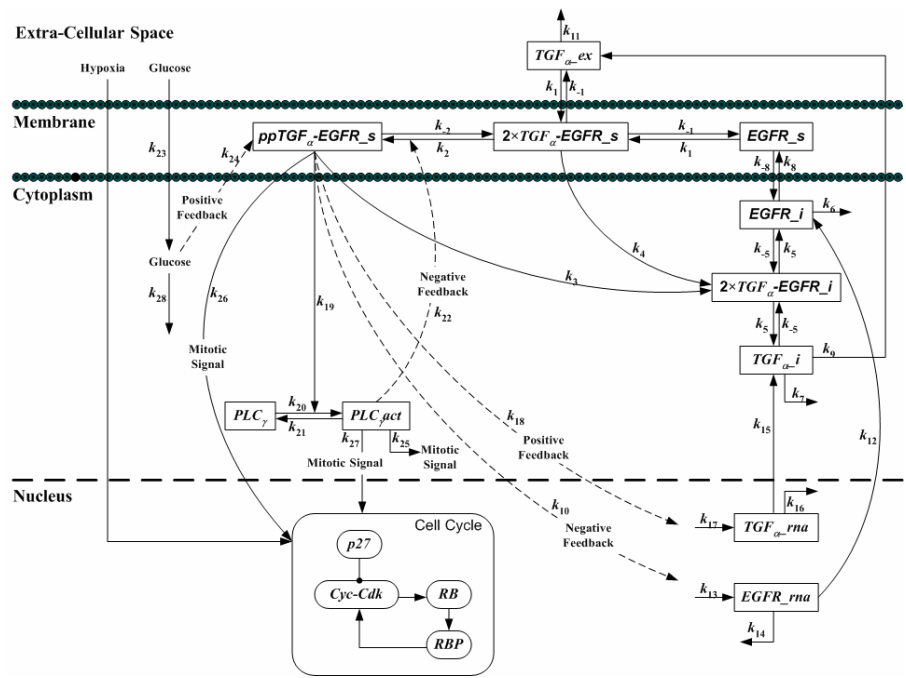

**FIGURE 4.**

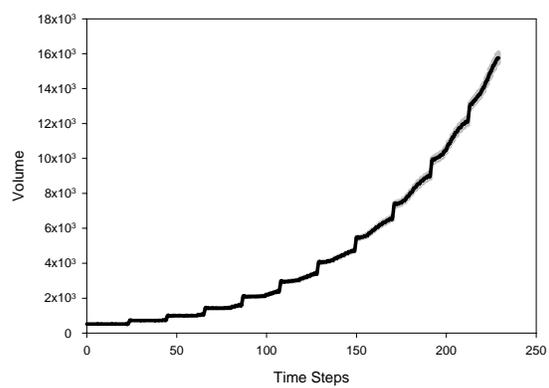

**FIGURE 5. (a)**

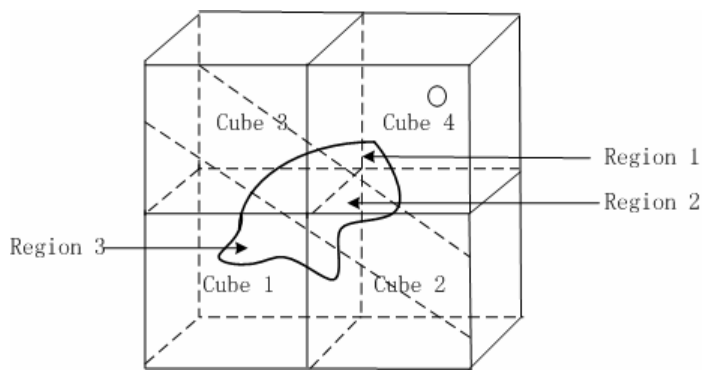

**FIGURE 5. (b)**

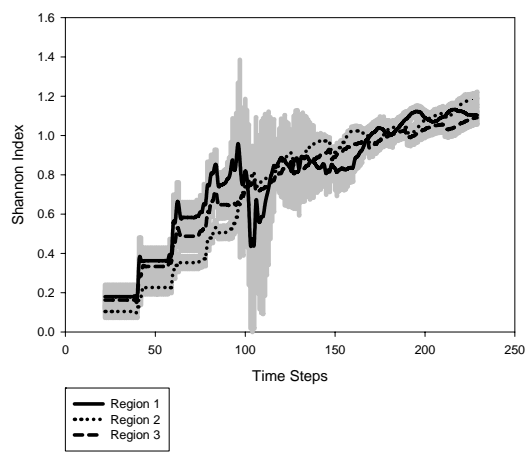

**FIGURE 6. (a)**

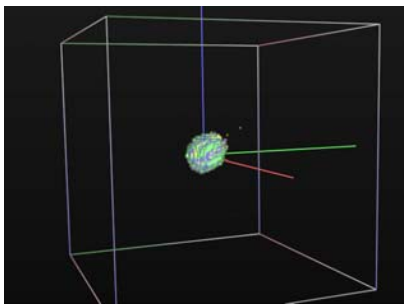

**FIGURE 6. (b)**

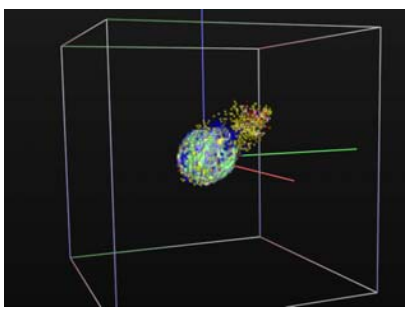

**FIGURE 6. (c)**

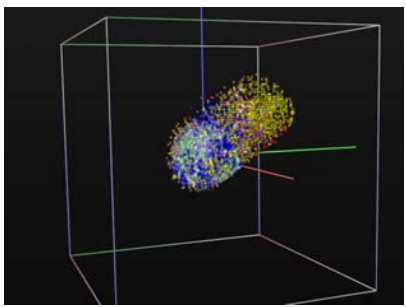

**FIGURE 7.**

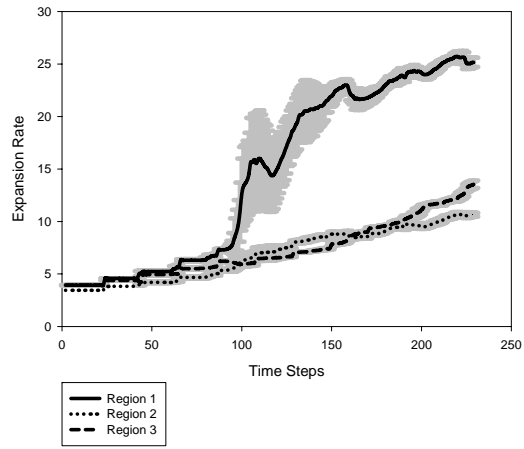

**FIGURE 8.**

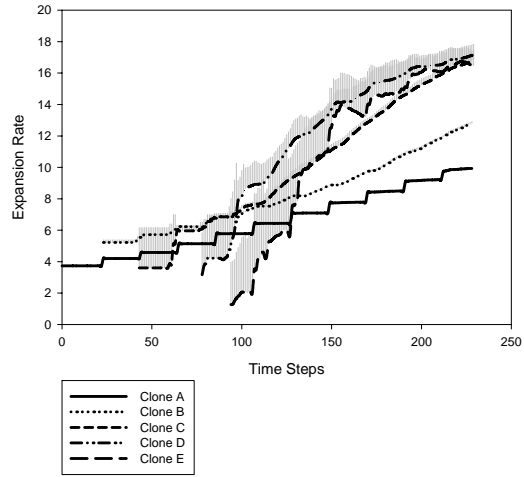

**FIGURE 9. (a)**

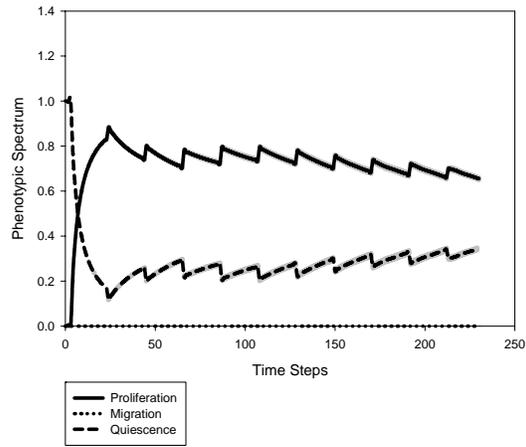

**FIGURE 9. (b)**

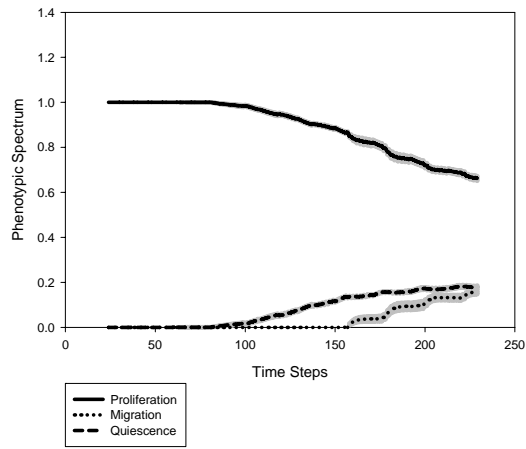

**FIGURE 9. (c)**

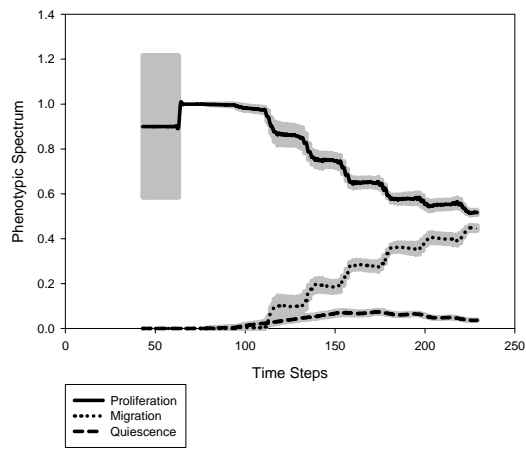

**FIGURE 9. (d)**

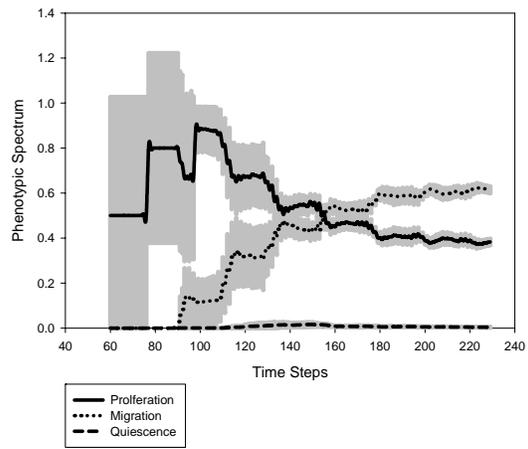

**FIGURE 9. (e)**

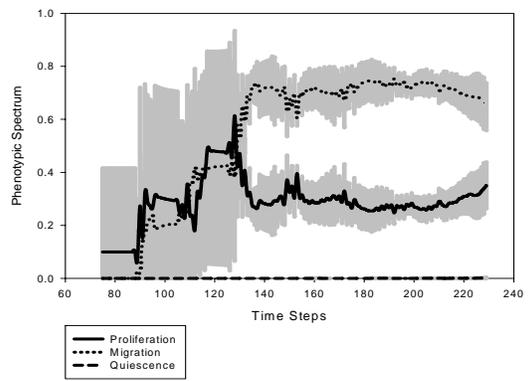

**Figure 10.**

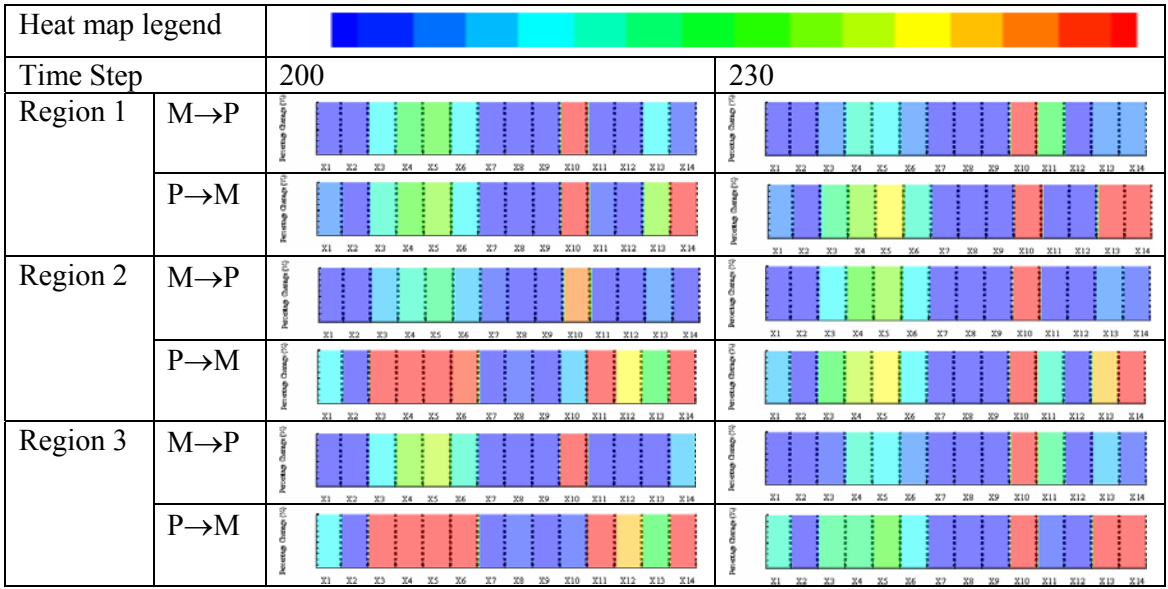

**TABLE 1.**

| Clone type | Cell cycle | Proliferation period |
|---|---|---|
| A | T | 2T |
| B | T | 1.8T |
| C | T | 1.6T |
| D | T | 1.4T |
| E | T | 1.2T |

**TABLE 2.**

| | | | |
|---|---|---|---|
| $r_A$ | $3.85 \times 10^{-12}$ | *mol/step* | Glucose uptake coefficient for clone A |
| $r_B$ | $4 \times 10^{-12}$ | *mol/step* | Glucose uptake coefficient for clone B |
| $r_C$ | $7.7 \times 10^{-12}$ | *mol/step* | Glucose uptake coefficient for clone C |
| $r_D$ | $1.54 \times 10^{-11}$ | *mol/step* | Glucose uptake coefficient for clone D |
| $r_E$ | $3.08 \times 10^{-11}$ | *mol/step* | Glucose uptake coefficient for clone E |
| $D_1$ | $6.7 \times 10^{-7}$ | $cm^2 s^{-1}$ | Diffusion coefficient of glucose |
| $D_2$ | $5.18 \times 10^{-7}$ | $cm^2 s^{-1}$ | Diffusion coefficient of $TGF_\alpha$ |
| $D_o$ | $8 \times 10^{-5}$ | $cm^2 s^{-1}$ | Diffusion coefficient of oxygen tension |
| $T_m$ | $147 \pm 18$ | $pg / ml$ | Maximum concentration of $TGF_\alpha$ |
| $G_a$ | $0.72 \times 10^{-9}$ | *mol/site* | Normal concentration of glucose |
| $G_m$ | $2.36 \times 10^{-9}$ | *mol/site* | Maximum concentration of glucose |
| $S_T$ | 0.3 | *molecules/min* | Secretion rate of $TGF_\alpha$ |
| $k_a$ | 0.0017 | *Dimensionless Constant (DC)* | Normal concentration of oxygen |
| $k_m$ | 0.0025 | *DC* | Maximum concentration of oxygen |

**TABLE 3.**

| Symbol | Variable |
|---|---|
| $X_1$ | $TGF\alpha$ extracellular protein |
| $X_2$ | $EGFR$ cell surface receptor |
| $X_3$ | Dimeric $TGF\alpha - EGFR$ cell surface complex |
| $X_4$ | Phosphorylated active dimeric $TGF\alpha - EGFR$ cell surface complex |
| $X_5$ | Cytoplasmic inactive dimeric $TGF\alpha - EGFR$ complex |
| $X_6$ | Cytoplasmic $EGFR$ protein |
| $X_7$ | Cytoplasmic $TGF\alpha$ protein |
| $X_8$ | $EGFR$ $RNA$ |
| $X_9$ | $TGF\alpha$ $RNA$ |
| $X_{10}$ | $PLC_\gamma$ inactive , Ca-bound |
| $X_{11}$ | $PLC_\gamma$ active, phosphorylated, Ca-bound |
| $X_{12}$ | Nucleotide pool |
| $X_{13}$ | Glucose cytoplasmic |
| $X_{14}$ | Glucose extracellular |
| $X_{15}$ | cdh1-APC complex |
| $X_{16}$ | cyclin-CDK |
| $X_{17}$ | Mass of the cell |
| $X_{18}$ | Protein p27 |
| $X_{19}$ | RBNP |